# Picosecond imaging of dynamics of solvated electrons during femtosecond laser-induced plasma generation in water

Noritaka Sakakibara,[1,2,3,*] Tsuyohito Ito,[1] Yukiya Hakuta,[2] Yoshiki Shimizu,[2] Kazuo Terashima,[1,2] and Eisuke Miura[2,*]

[1] Department of Advanced Materials Science, Graduate School of Frontier Sciences, The University of Tokyo, 5-1-5 Kashiwanoha, Kashiwa, Chiba 277-8561, Japan

[2] AIST-UTokyo Advanced Operando-Measurement Technology Open Innovation Laboratory (OPERANDO-OIL), National Institute of Advanced Industrial Science and Technology (AIST), 5-1-5 Kashiwanoha, Kashiwa, Chiba, 277-8589, Japan

[3] Department of Chemistry, School of Science, Tokyo Institute of Technology, 2-12-1-NE-2 Ookayama, Meguro, Tokyo 152-8550, Japan

[*] E-mail: nori.sakakibara@gmail.com

[*] E-mail: e-miura@aist.go.jp

**Abstract:** The dynamics of solvated electrons were visualized using absorption imaging with sub-picosecond time resolution based on a pump-probe measurement during the early stages of femtosecond laser-induced plasma generation in water. The solvated electrons were generated by the propagation of a femtosecond laser pump pulse. In the area with a pump laser intensity over $2\times10^{13}$ W/cm$^2$, where a high density of free electrons was produced, solvated electrons exhibited an additional rapid increase in optical density (OD) at 800 nm, 7–9 ps after the pump pulse excitation. In contrast, no two-step increase in OD was observed when probed at 400 nm, suggesting that the absorption coefficient of the solvated electrons rapidly changed around 800 nm after femtosecond laser excitation for a few picoseconds. This observation might indicate the structural and electronic modulation of solvated electrons owing to the high density of free electrons in water, accompanied by femtosecond-laser-induced plasma generation.

# I. INTRODUCTION

Femtosecond-laser-induced plasma generated in aqueous media is attracting attention for applications such as nanomaterials design,[1-4] cellular nanosurgery,[5-7] surface modification,[8] and carbon dioxide fixation.[9] By tightly focusing an intense femtosecond laser pulse with an energy of 1 mJ or more,[10] a high-density plasma with $10^{18}$–$10^{21}$ cm$^{-3}$ electron density is induced as a result of multiphoton ionization, tunneling ionization, and collisional ionization.[11, 12] The abundant free electrons are simultaneously hydrated by water molecules to form solvated electrons ($e_{aq}^-$).[13-15] The solvated electrons, with high reduction power (−3.1 V vs the standard hydrogen electrode),[16, 17] highly influences chemical reactions[18-20] and physical processes[11, 21] in aqueous media. In the field of pulse radiolysis and pulse photolysis research, the physical and chemical properties of solvated electrons have been thoroughly examined. In the photolysis studies, a low energy (10–100 μJ) femtosecond laser pulse is used. [19, 22-27] When focusing an intense femtosecond laser pulse (> 1 mJ), in contrast, the dynamics of solvated electrons can be influenced by the abundant free electrons provided by the laser-induced plasma. However, there have been no investigations into the dynamics of solvated electrons during plasma generation using an intense femtosecond laser pulse. It is also noteworthy that experimental investigations of solvated electrons in the plasma environment in an aqueous solution are very limited,[28-31] although the significance of solvated electrons in plasma chemistry has been reported in the recent decade.[32-35]

Recently, our group observed the decay dynamics of solvated electrons during plasma generation in water using femtosecond laser pulses.[36] The solvated electrons in the plasma environment showed a prolonged apparent lifetime of geminate recombination in a 100 ps–1 ns time scale, indicative of the influence of high-density free electrons in laser-induced plasma on the dynamics of solvated electrons. On the other hand, the influence of laser-induced plasma on the production dynamics of solvated electrons before the decay of solvated electrons, that is, on a picosecond timescale, has not been investigated.

In this study, we observed the dynamics of solvated electrons on the picosecond timescale during the focusing of an intense femtosecond laser pulse into water, by constructing a pump-probe absorption imaging system with sub-picosecond time resolution. Solvated electrons were generated according to the propagation of a femtosecond pump laser pulse. Subsequently, solvated electrons exhibited a further increase in optical density (OD) when using an 800 nm probe pulse with a 7–9 ps delay after the pump pulse propagation in the area where the pump laser intensity was higher than $2\times10^{13}$ W/cm$^2$ and a high density of free electrons was produced. In contrast, a two-step increase in optical density was not observed when probed at 400 nm, suggesting that the absorption coefficient of the solvated electrons changed at approximately 800 nm after femtosecond laser excitation for a few picoseconds. This observation might indicate a presumed structural and electronic modulation of solvated electrons due to the high density of free electrons in water, accompanied by femtosecond-laser-induced plasma generation.

## II. EXPERIMENT

**Observation of dynamics of solvated electrons**

Figure 1 shows pump-probe measurement setup. Laser pulses with a pulse width of 50 fs and center wavelength of 800 nm were delivered from a Ti:sapphire laser system using chirped-pulse amplification. The laser system consists of a mode-locked oscillator, pulse stretcher, regenerative amplifier, multipass amplifier, and pulse compressor. The laser system operated at a repetition rate of 1 Hz. One pulse, which was used as a pump pulse with a diameter of 10 mm, was focused using an aspherical quartz lens with 12.5 mm of focal length into the middle of a quartz cell filled (inner size: 10×10×45 mm) with purified water (electrical conductivity of 0.07 μS/cm). The pump-pulse induced the generation of plasma in water. The pump pulse energy was regulated at 0.3–10 mJ before it entered a focusing lens. The other pulse, which had a diameter of 6 mm and was used as the probe pulse, illuminated the plasma in a direction perpendicular to that of the pump pulse propagation. The energy of the probe pulse is < 0.2 mJ. Time-resolved observations were performed by controlling the optical delay between the pump and probe pulses by changing the optical length of the probe pulse using a micrometer. As will be shown later, the time resolution of our method was confirmed to be better than 1 ps.

Using the pump-probe technique, we constructed a picosecond time-resolved imaging system for absorption and interferometry measurements. The shadowgraph images were recorded using a charge coupled device (CCD) camera through a lens with a focal length of 150 mm. A probe pulse with a center wavelength of 800 nm was used for shadowgraph imaging because the wavelength is close to the peak wavelength (720 nm) of the absorption spectrum of solvated electrons.[37] OD was calculated as the logarithm of the ratio of pixel intensities of shadowgraph images with and without pump pulses at each position, as shown in Eq. (1), and the distribution of absorption by solvated electrons was visualized.

$$OD(x,y) = -log_{10}\left(\frac{I_{ON}(x,y)}{I_{OFF}(x,y)}\right), \quad (1)$$

where $x$ is the position in the direction of the pump pulse propagation, y is the position in the direction vertical to $x$, $I_{ON}$ is the pixel intensity of the shadowgraph image with a pump pulse, and $I_{OFF}$ is that without a pump pulse.

Interferometry images were obtained by inserting a fused silica biprism (with a top angle of 177°)[38] behind the focusing lens of a CCD camera to estimate the density of free electrons. For interferometry imaging, a probe pulse of 400 nm was used to avoid severe absorption of the probe pulse by the solvated electrons. The wavelength of the probe pulse was converted from 800 to 400 nm by second-harmonic generation using a 0.2-mm-thick beta barium borate (BBO) crystal. Behind the BBO crystal, we inserted two beam splitters for harmonic separation with an anti-reflection coating (CVI, BSR-48-UV 400R/800T) to remove the residual 800 nm-beam, although these optics were not shown in Figure 1. A probe pulse of 400 nm was also used for the shadowgraph imaging measurements. All the shadowgraphs and interferometric images were captured using a single laser shot.

**Calculation of density of solvated electrons**

The OD values are described by Eq. (2) based on the Lambert-Beer law:

$$OD(x) = \int_{-y_0}^{y_0} \varepsilon n_{e_{aq}^-}(r)dy = \int_0^{y_0} \frac{2\varepsilon n_{e_{aq}^-}(r)dr}{(r^2-x^2)^{\frac{1}{2}}} \quad (2)$$

where $x$ is the direction of pump pulse propagation, y is the vertical direction to $x$, $r$ is the radial direction of the pump pulse, $\varepsilon$ is the absorption coefficient of solvated electrons at the wavelength of probe pulse (18920 M$^{-1}$cm$^{-1}$ for 800 nm and 3350 M$^{-1}$cm$^{-1}$ for 400 nm),[37] $n_{e_{aq}^-}$ is the density of solvated electrons. Applying the inverse Abel transformation[39] to Eq. (2), the density distribution of solvated electrons was obtained using Eq. (3).

$$n_{e_{aq}^-}(r) = -\frac{1}{2\pi\varepsilon}\int_r^{r_0}\frac{OD(x)dx}{(x^2-r^2)^{\frac{1}{2}}} \qquad (3)$$

Inverse Abel transformation was applied to the images of OD values with the laser path of a pump pulse as axisymmetric using the freely available code of MATLAB[40] under the assumption of an axisymmetric distribution of solvated electrons.

**Calculation of the density of free electrons**

Interferometry images were obtained with a 5 mJ pump pulse or without the pump pulse, using the same pump-probe system for the OD imaging measurement. Phase shift component ($\delta\varphi$) was deduced from the interferometry image by Fourier transformation and subsequently inverse Abel transformation was applied to $\delta\varphi$ with the laser path of a pump pulse as axisymmetric, resulting in the left term of Equ. (4). The density of free electrons ($n_e$) was calculated using the Drude model, as described by Eq. (4).

$$-\frac{1}{2\varepsilon\pi}\int_r^{r_0}\frac{\delta\varphi dx}{(x^2-r^2)^{\frac{1}{2}}} = -\frac{e^2\lambda}{2\pi\varepsilon_r\varepsilon_0 m_e c^2} \times n_e \qquad (4)$$

where $m_e$ is the mass of an electron, $e$ is the elementary charge, $\lambda$ is the wavelength of the probe pulse, $\varepsilon_r$ is the relative permittivity of water, $\varepsilon_0$ is the permittivity of vacuum and $c$ is the speed of light. All processing was performed by constructing a MATLAB code under the assumption of an axisymmetric distribution of free electrons.

**Calculation of ionization rate and Keldysh parameter in water**

Ionization rates were calculated in the same manner as in a previous report by Efimenko et al.[41] Ionization rates with strong electromagnetic field (i.e., tunneling ionization and multiphoton ionization) were calculated based on Keldysh theory using the following parameters; 800 nm of pump pulse wavelength, 6.5 eV of ionization potential for water molecules, 1.33 of dielectric constant of water molecules and $3.33\times10^{22}$ cm$^{-3}$ of water molecules density. The Keldysh parameter, which is an indicator of the multiphoton ionization and tunneling ionization regimes, was also calculated for an 800 nm pump pulse propagating in water as a guide to recognizing the regime of the laser-induced plasma. The rate of collisional ionization was calculated based on Drude's theory.[41, 42]

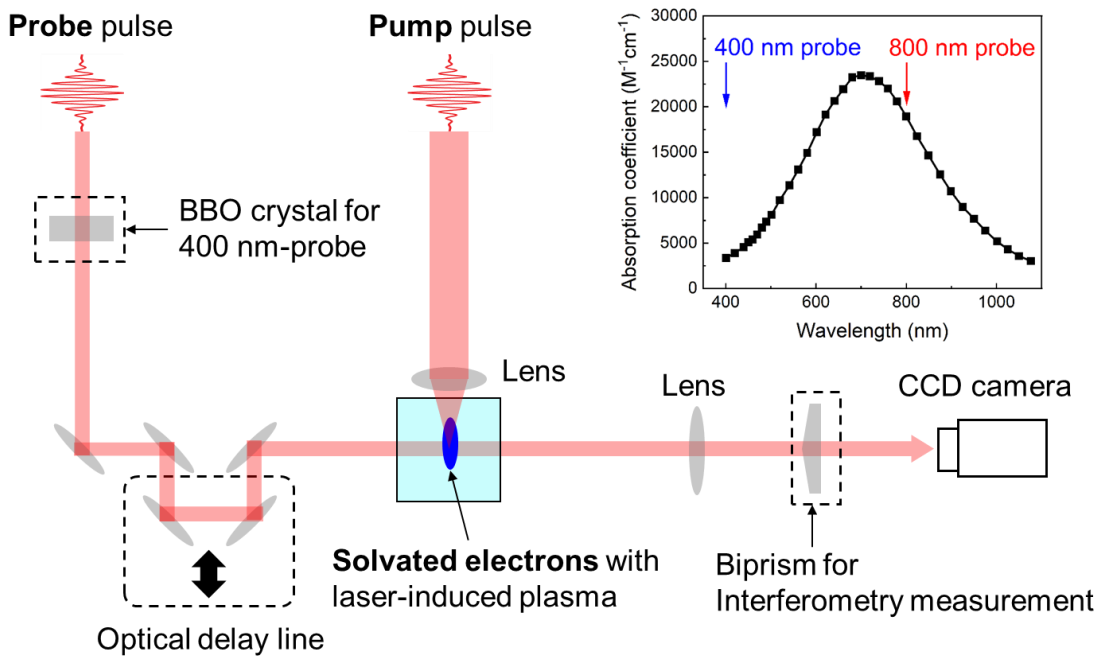

**Figure 1.** Schematic illustration of the pump-probe system for the picosecond imaging measurements of solvated electrons and free electrons during the generation of laser-induced plasma in water. Pump pulse: pulse energy 0.3–10 mJ, pulse width 50 fs, wavelength 800 nm. Probe pulse: pulse energy < 0.2 mJ, pulse width 50 fs, wavelength 800 nm. BBO crystal was inserted when probe pulse with the wavelength of 400 nm was used. Biprism was inserted when the interferometry measurement was performed. The absorption coefficient spectrum of solvated electrons is shown in the up-right corner of the figure with the wavelengths of probe pulses used in this study. The data of the absorption coefficient is adopted from the reference.[37]

## III. RESULTS

### Imaging of solvated electrons using 800 nm probe pulse

During the propagation of the pump pulse by focusing it in water in a cuvette cell, shadowgraph images were captured at a picosecond time resolution by changing the probe delays using the optical delay line. The shadowgraph images were converted to OD images by calculating the logarithm of the ratio of the pixel intensities of the shadowgraph images with and without pump pulses. Figure 2(a) and Movie 1 show picosecond time-resolved images of the OD distribution monitored by an 800 nm probe pulse during the propagation of the pump pulse with 5 mJ pulse energy. In the images, a pump pulse was propagated and focused from the right side of the images along the $x$ direction. The time at which the shadow of the pump pulse reached the focal point was defined as a probe delay of $t = 0$ ps. Figure 2(b) shows the time evolution of the OD values at the lateral ($y$-directional) center of the OD distribution at different axial ($x$) positions. A sufficient change in the OD value was observed with a delay time change of 1 ps as shown in Figures 2(a) and 2(b), confirming that the proposed method had a time resolution of at least 1 ps.

The OD values increased along the pump laser path according to the pump pulse propagation at $t < 0$ ps and formed a distribution shape similar to a focusing cone at $t = 0$ ps. The OD is attributed to the absorption of solvated electrons, which was confirmed by a scavenger experiment with $NO_3^-$ in our previous study.[36] The larger OD values closer to the focal point indicate a higher yield of solvated electrons owing to the higher pump pulse intensity by the focusing. The laser intensity, calculated from the pump pulse energy and beam diameter (full width at half maximum of the OD distribution in the $y$ direction), was greater than $2 \times 10^{13}$ W/cm$^2$ near the focal point ($x < 4$ mm). The laser intensity exceeded the breakdown threshold reported previously for femtosecond laser-induced breakdown in water.[10, 43] For the calculation of the laser intensity, we did not consider the pump pulse energy loss during the propagation because the absorption of the laser by plasma, which should be a main loss of pump pulse energy, can be ignored as the free electron density (as described later) was significantly lower than the cut-off density at 800 nm ($1.7 \times 10^{21}$ cm$^{-3}$) and the pump pulse energy loss should be small. The OD values increased sharply within a few picoseconds (within 1 ps at $x = 3$ mm, Figure 2(b)) near the focal point, whereas a gradual increase in the OD values was observed (~5 ps at $x = 5$ mm, Figure 2(b)) in the focusing cone far from the focal point ($x >$ 4 mm).

As shown in Figure S1, the ionization rate of water depends on the laser intensity and is higher at higher laser intensities. When the laser intensity was greater than $2\times10^{13}$ W/cm$^2$, the ionization time, which is the reciprocal of the ionization rate, was approximately 1 ps or less (Figure S1(b)). Considering the ionization time is close to or shorter than the laser pulse duration (50 fs) with Keldysh parameter $\gamma < 1$, tunneling ionization should be the dominant ionization mechanism. In contrast, when the laser intensity is lower than $10^{13}$ W/cm$^2$, ionization progressed gradually with an ionization time of approximately 10 ps or longer, which could be dominated by collisional ionization and multiphoton ionization. The time evolution of the OD values described above (Figure 2) should reflect the intensity-dependent ionization behavior of water. Although the laser intensity was high especially near the focal point, non-linear phenomena such as self-focusing was not observed in this study.

After a pump pulse passed through the focal point ($t > 0$ ps), a second steep evolution of the OD was observed 7–9 ps after the first OD evolution near the focal point ($x < 4$ mm). The evolved OD values remained constant until $t = 20$ ps. The two-step evolution of the OD values is clearly observed in the time evolution of the OD values at the lateral center of the OD distribution at different $x$ positions (Figure 2(b)). When the $x$ position was closer to the focal point, the slope of the second OD evolution became steeper, clearly exhibiting a two-step evolution of the OD values. In contrast, the two-step evolution of the OD values became less evident ($x = 4.5$ mm) with the gradual slope of the second OD increase and was not observed clearly ($x > 5$ mm) away from the focusing cone. The $x$-position dependence of the slope of second OD increase showed the same position dependence as that of the first OD increase (Figure S2), which can be explained by the change in ionization behavior according to the focusing of pump pulse as described above. On the other hand, the duration of the plateau between the first and second OD rises became longer at closer to the focal point (Figure S2(a)). As such, the total time for the plateau and second OD rise, *i.e.*, the time between the two OD increases, was similar (8 or 9 ps) at any $x$-position (Figure S2(a)). Figure 3(b) demonstrates the distribution of a difference in OD values (ΔOD) between time delays of 0 and 10 ps, exhibiting the area that appeared the two-step OD evolution. The green dashed line indicates the boundary where a two-step evolution of OD values was observed based on the ΔOD image (Figure 3(b)). There is a threshold laser intensity of approximately $2\times10^{13}$ W/cm$^2$ for the two-step OD evolution. These observations imply that the dynamics of solvated electrons are influenced even after a pump pulse passes through the focal point and that the extent of the influence strongly depends on the pump pulse laser intensity.

Finally, the OD values gradually decreased on a timescale of a few nanoseconds, and a small plasma plume grew in the vicinity of the focusing cone at $x < 3.6$ mm (Figure S3(a)). The nanosecond decay kinetics of OD values, which were different from the picosecond decay kinetics of solvated electrons owing to geminate recombination, as often observed in photolysis studies,[22, 25] were fitted well by a modified geminate recombination model considering the additional production of solvated electrons (Figures S3(b) and S3(c)), as already investigated in our previous study.[36] Cavitation bubbles were not observed up to $t = 2.7$ ns (maximum time delay in this study) because the time scale of the formation and growth of cavitation bubbles are in nanoseconds to microseconds timescale,[44-46] indicating that the influence of cavitation bubbles on the shdaowgraph and interferometry measurements should be quite small.

The OD distribution images were converted using an inverse Abel transformation under the assumption of an axisymmetric distribution of solvated electrons, and the distribution of the density of solvated electrons was evaluated using Eq. (3). An image of the distribution of the solvated electrons at $t = 20$ ps is shown in Figure S4(b). The time evolutions of the radial distributions of the solvated electrons at $x = 3$ and 4.5 mm are shown in Figures S4(c) and S4(d), respectively. The observation of the distribution of solvated electrons with sub-picosecond time resolution is an advantage of the imaging method constructed in this study. The maximum density of solvated electrons at $t = 20$ ps was calculated to be $4.2\times10^{17}$ cm$^{-3}$. This value is consistent with our previous study, in which the area-averaged density of solvated electrons was evaluated using dual-probe beams that passed through the pump region or not,[36] therefore, the validity of our picosecond imaging method was confirmed. It should be noted that the calculated density of the solvated electrons near the focal point (especially before $t = 10$ ps) was tentative because the absorption coefficient of the solvated electrons might be affected, as discussed later.

The dynamics of solvated electrons were also observed with different pump pulse energies using an 800 nm probe pulse. As shown in Figure 4, the area of the shadow due to solvated electron generation became narrower as the pump pulse energy decreased from 10 to 0.34 mJ. The area where the two-step dynamics of the solvated electrons were observed (Figure S5) also became smaller. In contrast, the threshold intensity of the pump pulse necessary for the two-step dynamics was $1.8\times10^{13}$–$3.1\times10^{13}$ W/cm$^2$, which was maintained at different pump pulse energies of approximately $2\times10^{13}$ W/cm$^2$. Thus, the two-step dynamics of the solvated electrons is a ubiquitous phenomenon when the pump pulse exceeds the threshold intensity. Temporal OD evolutions with 1.2 and 0.34 mJ pump pulses (Figure S6) also supported the two-step dynamics of solvated electrons. With the pump pulses, the time between the first and second OD increases was 7-9 ps (Figures S6(b) and S6(d)), indicating that the two-step dynamics is not dependent on the pulse energy but on the focusing intensity of the pump pulse.

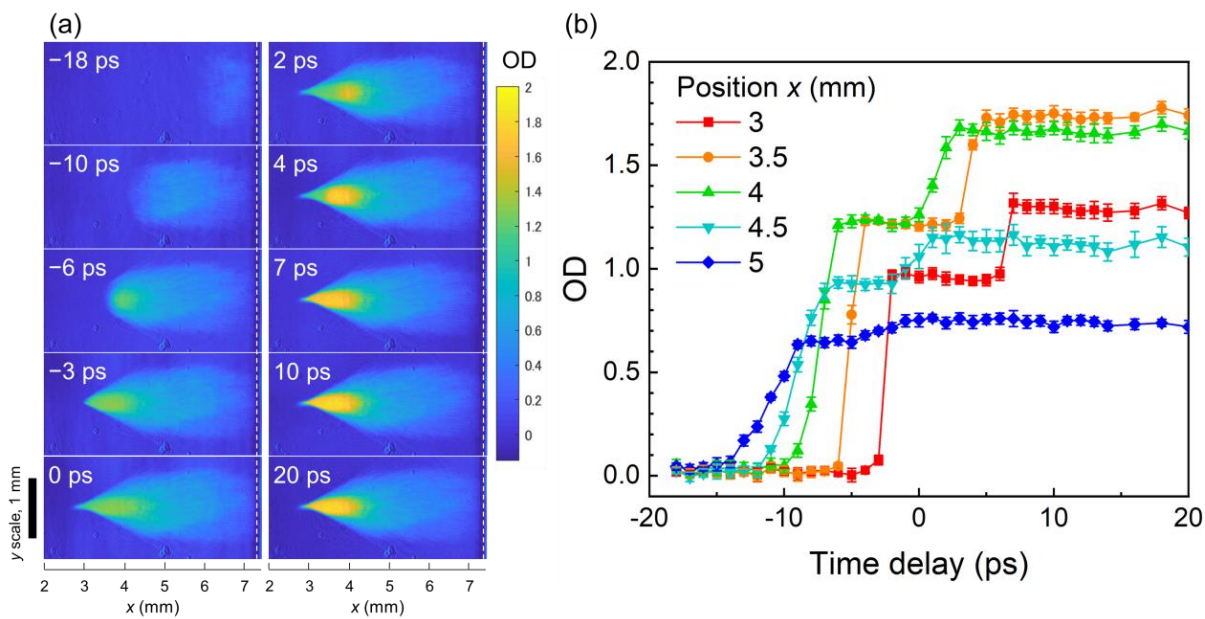

**Figure 2.** Dynamics of solvated electrons probed at 800 nm. (a) OD distributions monitored at different time delays during the propagation of the pump pulse (5 mJ). (b) Time evolution of OD at the lateral (*y*-directional) center of the OD distribution with different axial (*x*) positions. The pump pulse propagated from the right side of the images. The white dashed lines at the right side of the images indicate the wall of a cuvette. A movie of the temporal OD distribution with 1 ps time resolution is available in Movie 1.

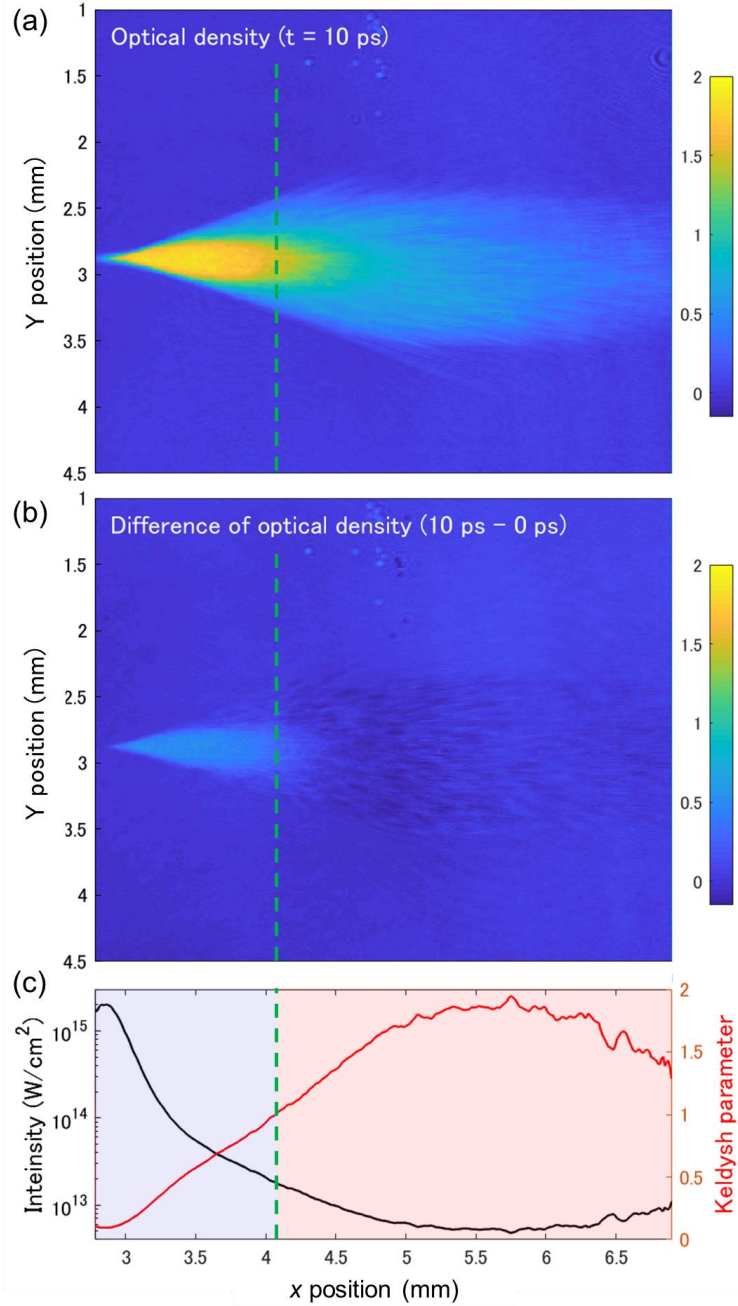

**Figure 3.** (a) OD image at $t$ = 10 ps, (b) image of a difference of OD values ($\Delta$OD) between $t$ = 0 ps and 10 ps. The OD images were obtained during the propagation of the pump pulse (5 mJ) probed at 800 nm. The color bar indicates the value of OD. (c) Distributions of pump pulse laser intensity and corresponding Keldysh parameter along the $x$ direction at the lateral center of the OD distribution. The green line indicates the boundary where a two-step evolution of OD values was observed based on $\Delta$OD in Figure 3(b). The two-step evolution occurred where the laser intensity was higher than approximately $2\times10^{13}$ W/cm$^2$.

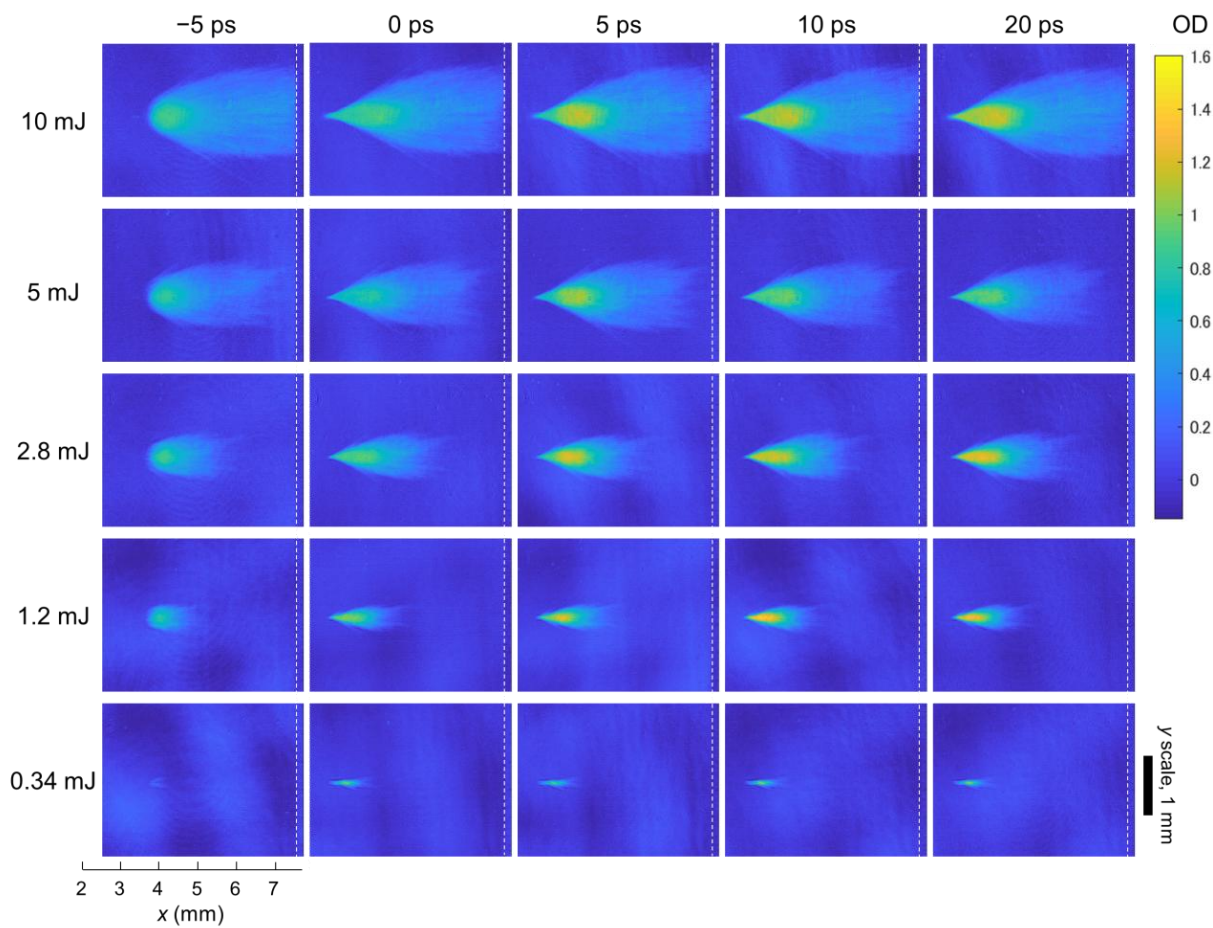

**Figure 4.** Temporal OD distribution probed at 800 nm with changing the pump pulse energy from 0.34 to 10 mJ. From the images, the threshold intensity required for the occurrence of a two-step evolution of OD values was estimated at each pump pulse energy (Figure S5). In Figure S5, ΔOD images at $t = 0$, 5, 10 and 20 ps between $t = 0$ ps with different pulse energies are shown. The color bar indicates the value of OD. The pump pulse propagated from the right side of the images. The white dashed lines at the right side of the images indicate the wall of a cuvette.

**Imaging of free electrons and imaging of solvated electrons using 400 nm probe pulse**

As described above, two-step steep OD evolution was observed near the focal point ($x < 4$ mm) with laser intensity over approximately $2\times10^{13}$ W/cm$^2$. With such a high intensity pump pulse, tunneling ionization governs the generation of electrons in a femtosecond timescale (Figure S1) and generates a high density of solvated electrons up to $4.2\times10^{17}$ cm$^{-3}$ (Figure S4), suggesting that high density of free electrons should be generated.

Interferometry images were obtained at picosecond time resolution by changing the probe delays (Figure S7). The distribution of free electrons was calculated using the Drude model, as described in Eq. (4). The distribution of the density of free electrons at different probe delays is shown in Figure 5(a), and the temporal behavior of the density of free electrons is shown in Figure 5(b). Near the focal point ($x = 3.2$ mm), the density of free electrons increased steeply between $t = -2$ and 0 ps up to $1.1\times10^{19}$ cm$^{-3}$ by the focusing of a pump pulse. The density of free electrons was higher than that of solvated electrons by more than one order of magnitude, which is consistent with our previous study in which an 800 nm probe was used.[36] Since a part of free electrons forms solvated electrons, it is reasonable that the density of solvated electrons is much lower than that of free electrons. The density of free electrons decreased monotonically and became $9.2\times10^{18}$ cm$^{-3}$ at $t = 20$ ps. At $x = 3.7$ mm which is still within the area of two-step OD evolution, the density of free electrons increased steeply between $t = -4$ and $-2$ ps up to $3.1\times10^{18}$ cm$^{-3}$ and further increased until $t = 3$ ps up to $4.0\times10^{18}$ cm$^{-3}$. The density of free electrons was maintained until $t = 16$ ps. At $x = 4.2$ mm, which is outside of the area of two-step OD evolution, the density of free electrons increased gradually from $t = -8$ to 6 ps up to $0.9\times10^{18}$ cm$^{-3}$ and the density maintained until $t = 16$ ps. As such, the density of free electrons showed gradual decrease/increase or was maintained after the pump pulse passed away ($t > 0$ ps), in contrast to the second steep increase in the OD near the focal point.

When a probe pulse of 400 nm was used for shadowgraph imaging, as shown in Figure 6(a), the OD values evolved along the pump laser path according to the pump pulse propagation at $t < 0$ ps and formed a distribution shape similar to the focusing cone of the pump pulse at $t = 0$ ps, which is the same behavior as that monitored with an 800 nm probe pulse. The OD values of the 400 nm probe were significantly lower than those of the 800 nm probe owing to the lower absorption coefficient of the solvated electrons at 400 nm (Figure 1). After a pump pulse passed through the focal point ($t > 0$ ps), in contrast to the case of the 800 nm probe pulse, the OD values remained constant until $t = 20$ ps without further evolution of the OD value even around the focusing cone. Therefore, the two-step OD evolution that was observed with the 800 nm probe pulse was not observed with the 400 nm probe pulse. This behavior was also demonstrated in the temporal OD values at the lateral center of the OD distribution at different $x$ positions (Figure 6(b)). This result indicates that the two-step OD evolution of solvated electrons depends on the wavelength of the probe pulse.

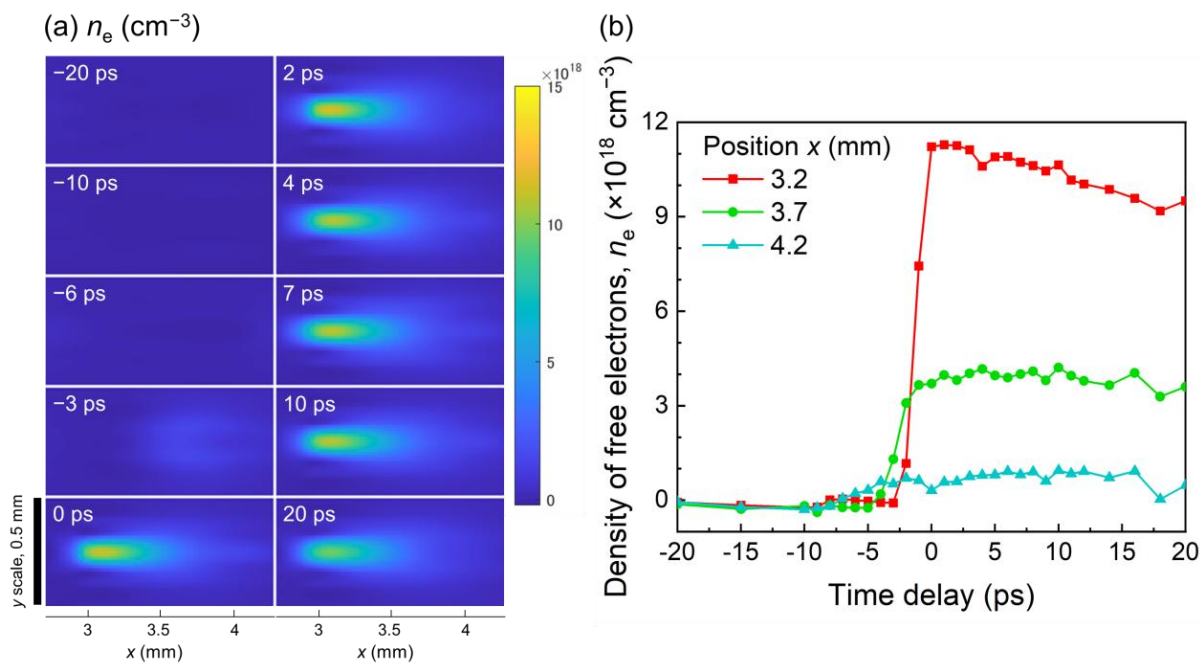

**Figure 5.** Dynamics of free electrons at 5 mJ of pump pulse energy estimated by interferometry images at different time delays. (a) Temporal images of the distribution of free electrons. (b) Temporal density of free electrons at the lateral center of the distribution image at $x = 3.2$, 3.7 and 4.2 mm. The interferometry images are shown in Figure S7.

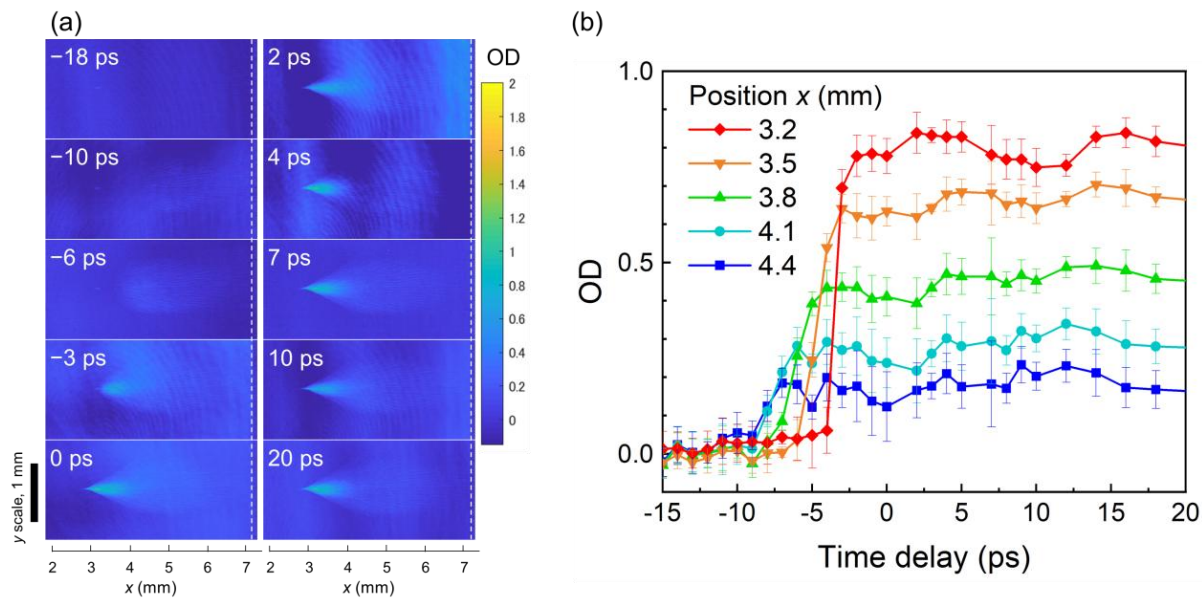

**Figure 6.** Dynamics of solvated electrons probed at 400 nm. (a) OD distributions monitored at different time delays during the propagation of the pump pulse (5 mJ). (b) Time evolution of OD at the lateral (*y*-directional) center of the OD distribution with different axial (*x*) positions. The pump pulse propagated from the right side of the images. The white dashed lines at the right side of the images indicate the wall of a cuvette. A movie of the temporal OD distribution with 1 ps time resolution is available in Movie 2.

## IV. DISCUSSION

To explain the two-step OD evolution of the solvated electrons, one possibility is to account for the additional formation of solvated electrons after the pump pulse passed away (t > 0 ps). The high density of free electrons (Figure 5) can be hydrated to form additional solvated electrons. However, the situation is not straightforward. The additional formation of solvated electrons cannot explain the maintenance of the OD values when probed at 400 nm (Figure 6) because such additional solvation would also increase the OD at 400 nm.

From the results shown in Figure 2(b) and Figure 6(b), the time evolution of the ratio of the OD value probed at 800 nm to that probed at 400 nm ($OD_{800nm} / OD_{400nm}$) was calculated at $x$ = 3.5, 4.0 and 4.5 mm (Figure 7). At the time delay of the first OD increase, $OD_{800nm} / OD_{400nm}$ was 1.9±0.1, 4.4±0.8 and 5.1±1.1 at 3.5, 4 and 4.5 mm, respectively. Because the optical path lengths of the probes were the same in both cases, the ratio corresponded to the ratio of the absorption coefficients between 800 and 400 nm. However, as shown in the inset of Figure 1, the ratio of the absorption coefficients of the solvated electrons at 800 and 400 nm is originally approximately 5.7. As shown in Figure 7, the ratio was significantly lower than 5.7 near the focal point (x = 3.5 mm) even after the second OD increase at 4 ps observed using an 800 nm probe. The ratio was slightly lower than 5.7 at the boundary position of the two-step OD increase (x= 4.0 mm) and became similar to 5.7 after the second OD increase at approximately 1 ps. In contrast, the ratio was approximately 5.7 (within the error bar) at $x$ = 4.5 mm, which was outside the area of the two-step OD evolution. Therefore, the shape of the absorption spectrum of solvated electrons presumably changed near the focal point ($x$ = 3.5 and 4.0 mm), where the density of free electrons was high and a two-step evolution of OD values was observed. The extent of the spectral shape change should be larger closer to the focal point. It should be noted that the maximum value of the density of solvated electrons at $t$ = 20 ps, which was calculated to be $4.2 \times 10^{17}$ cm$^{-3}$ in the former section based on the OD data at 800 nm probe using the absorption coefficient value of "normal" solvated electrons, was underestimated and should be corrected by considering the spectral change of the absorption of solvated electrons. The $OD_{800nm} / OD_{400nm}$ was 2.6 at $t$ = 20 ps, indicating that the solvated electrons near the focal point are still modulated. When comparing the $OD_{800nm}$ / $OD_{400nm}$ value (2.6) to that of standard solvated electron (5.7), the absorption coefficient at 800 nm at $t$ = 20 ps near the focal point should be 2.2 times smaller than that of standard electron. Therefore, the calculated solvated electron density ($4.2\times10^{17}$ cm$^{-3}$) based on the OD value was underestimated by 2.2 times and can be corrected to be $9.2\times10^{17}$ cm$^{-3}$. The corrected density of solvated electrons was still an order of magnitude lower than the free electron density, which is consistent with the fact that a part of free electrons forms solvated electrons. It is also noteworthy that similar spectral shape change of the absorption of solvated electrons has been recently reported on the surface of water cathode in contact with argon plasma, in which the ratio of the OD values probed at 700 nm to that probed at 450 nm was lower than that of "normal" solvated electrons.[47]

Near the focal point, there was a severe gradient of the density of free electrons with a high density of free electrons up to $1.1\times10^{19}$ cm$^{-3}$ (Figure 5(a)), which might have induced a strong electric field even after the pump pulse receded. The alignment of water molecules to ice-like structures under the influence of strong electric field (> 1 MV/cm) has been previously reported.[48] Ice-like tetrahedral structures induce spectral changes of the absorption of solvated electrons, especially near the wavelengths of absorption peak (700 nm), with blue-shifted absorption peak by suppressing translational motions of water molecules.[49] Stark effect by such a strong electric field might also cause a blue-shifted absorption peak of solvated electrons.[28] Consequently, we hypothesized that a locally-induced strong electric field due to the high density of free electrons might align water molecules and modulate the electronic state of solvated electrons, which changed the shape of absorption spectrum of solvated electrons. Assuming that the spectral change observed in our study largely decreases the absorption coefficient of solvated electrons at 800 nm, the release from the modulation to "normal" solvated electrons can cause an apparent increase of OD at 800 nm, corresponding to the second OD increase in the two-step OD evolution (Figure S8).

However, we could not find any experimental results that could explain the steep increase in the OD (within 1 ps at $x$ = 3 mm; Figure 2(b)) during the second OD increase. In order to elucidate the steep second OD increase and confirm the hypothesis of the modulated hydration state of solvated electrons by the high density of free electrons, further spectroscopic and kinetic studies and computational modeling are necessary. The dynamics of spectral change of the absorption of solvated electrons can be investigated using probes with several wavelengths between 800–400 nm and above 800 nm. Computational modeling[15, 50-52] such as density functional theory (DFT) and molecular dynamics (MD) simulation should be useful to investigate the electronic state of solvated electrons under high density of free electrons. Soft X-ray spectroscopy could be also useful for directly observing the bonding structure of water molecules.[53, 54] Although the origin of the two-step OD evolution is deferred for future investigation, our study suggests a plausible structural and electronic modulation of solvated electrons by the high density of free electrons in water, accompanied by femtosecond-laser-induced plasma generation.

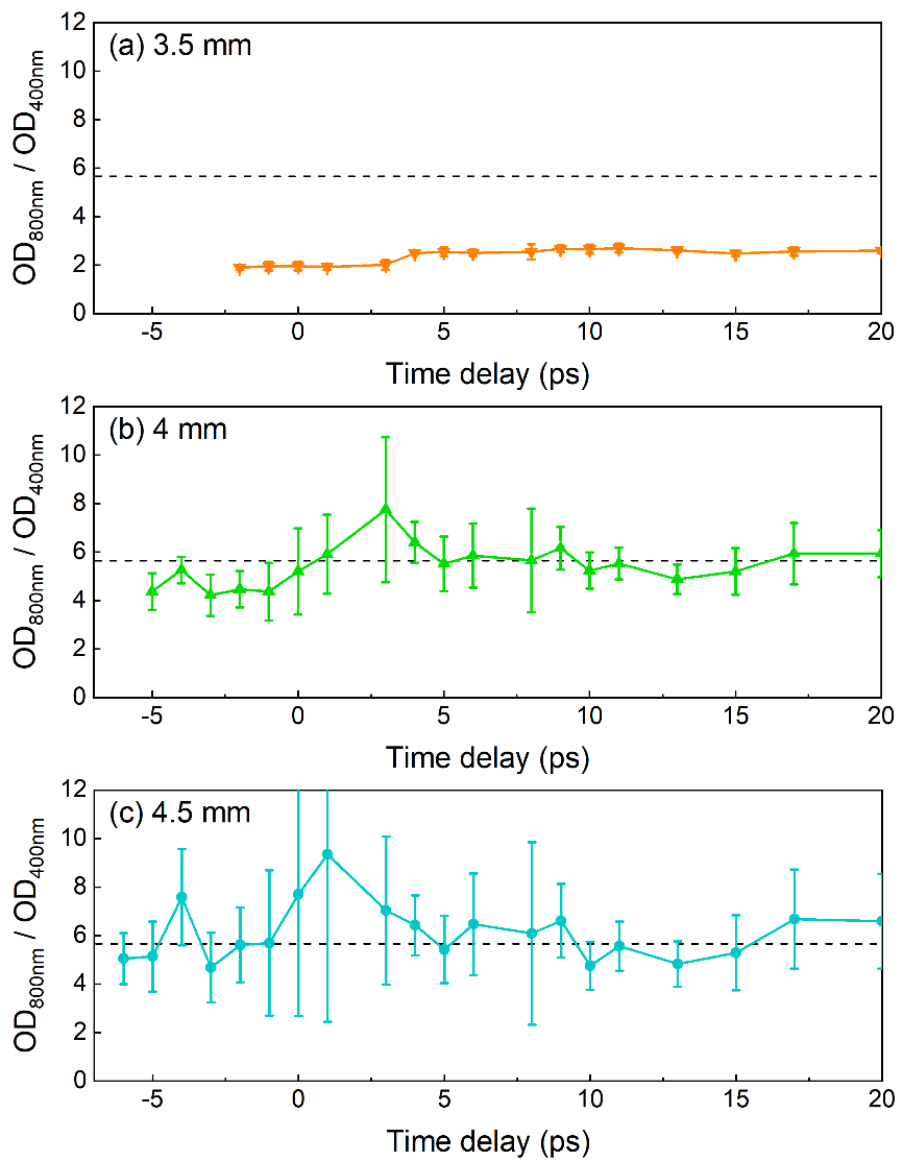

**Figure 7.** Time evolution of the ratio of OD values probed at 800 nm to that probed at 400 nm ($OD_{800nm}$ / $OD_{400nm}$) at (a) $x = 3.5$ mm, (b) $x = 4$ mm and (c) $x = 4.5$ mm. The OD values are obtained at the lateral center of the OD distribution with a 5 mJ pump pulse. The black dashed line indicates the ratio of the absorption coefficient of solvated electrons at 800 nm to that at 400 nm, which is 5.7, as calculated from the inset graph of Figure 1. The values of the ratio were plotted for the time delay after the first OD increase at each $x$ position.

## V. CONCLUSION

We observed the dynamics of solvated electrons on a picosecond timescale while focusing an intense femtosecond laser pulse into water by constructing a pump-probe imaging system with a sub-picosecond time resolution. Solvated electrons were generated according to the propagation of a femtosecond laser pump pulse. Subsequently, the solvated electrons exhibited a further rapid increase in the OD at 800 nm with a delay of 7–9 ps after the pump pulse excitation near the focal point, where the pump laser intensity was higher than $2\times10^{13}$ W/cm$^2$ and a high density of free electrons are generated up to $1.1\times10^{19}$ cm$^{-3}$. In contrast, the two-step increase in the OD was not observed when probed at 400 nm, suggesting a spectral shape change in the absorption of the solvated electrons. We hypothesized that the high density of free electrons and their large gradient, accompanied by femtosecond-laser-induced plasma generation, align water molecules and modulate the electronic state of solvated electrons. The structural and electronic modulation of solvated electrons might result in a two-step OD increase when probed at 800 nm by rapidly changing the absorption coefficient around 800 nm after femtosecond laser excitation for a few picoseconds, although further spectroscopic and kinetic studies are necessary to clarify this mechanism.

## SUPPLEMENTARY MATERIAL

The supplementary material contains laser intensity dependence of ionization rate, axial position dependence of times for the first and second OD increases, temporal dynamics of OD at picosecond to nanosecond timescales, radial distributions of OD and the density of solvated electrons, laser intensity dependence of a difference of ΔOD, laser intensity dependence of the temporal dynamics of OD, interferometry images and illustration of hypothesized spectral change of absorption coefficient of solvated electrons. The supplementary movies contain picosecond dynamics of OD probed at 800 nm (Movie 1) and 400 nm (Movie 2).

## AUTHOUR INFORMATION

**Corresponding author**

*Email: nori.sakakibara@gmail.com, e-miura@aist.go.jp

**Notes**

The authors declare no competing financial interest.

## ACKNOWLEDGEMENTS

N.S. thanks JSPS KAKENHI (Grant Numbers: JP19J13045, JP23K13821).

# REFERENCES

1. H. Belmouaddine, M. Shi, L. Sanche and D. Houde, *Phys. Chem. Chem. Phys.* **20** (36), 23403-23413 (2018).
2. L. M. Frias Batista, V. K. Meader, K. Romero, K. Kunzler, F. Kabir, A. Bullock and K. M. Tibbetts, *J. Phys. Chem. B* **123** (33), 7204-7213 (2019).
3. V. K. Meader, M. G. John, C. J. Rodrigues and K. M. Tibbetts, *J. Phys. Chem. A* **121** (36), 6742-6754 (2017).
4. A. A. Astafiev, A. M. Shakhov, A. G. Tskhovrebov, A. A. Shatov, M. S. Syrchina, D. V. Shepel and V. A. Nadtochenko, *ACS Appl. Nano Mater.* (2024).
5. V. Venugopalan, A. Guerra, K. Nahen and A. Vogel, *Phys. Rev. Lett.* **88** (7), 078103 (2002).
6. A. Vogel, J. Noack, G. Hüttman and G. Paltauf, *Appl. Phys. B* **81** (8), 1015-1047 (2005).
7. P. Ronchi, S. Terjung and R. Pepperkok, *Biol. Chem.* **393** (4), 235-248 (2012).
8. H. Muneoka, T. Koike, T. Ito, K. Terashima and E. Miura, *J. Phys. D: Appl. Phys.* **57** (24), 245205 (2024).
9. V. V. Rybkin, *J. Phys. Chem. B* **124** (46), 10435-10441 (2020).
10. C. H. Fan, J. Sun and J. P. Longtin, *J. App. Phys.* **91** (4), 2530-2536 (2002).
11. N. Linz, S. Freidank, X.-X. Liang and A. Vogel, *Phys. Rev. B* **94** (2), 024113 (2016).
12. S. S. Mao, F. Quéré, S. Guizard, X. Mao, R. E. Russo, G. Petite and P. Martin, *Appl. Phys. A* **79** (7), 1695-1709 (2004).
13. B. Abel, U. Buck, A. L. Sobolewski and W. Domcke, *Phys. Chem. Chem. Phys.* **14** (1), 22-34 (2012).
14. K. R. Siefermann, Y. Liu, E. Lugovoy, O. Link, M. Faubel, U. Buck, B. Winter and B. Abel, *Nat. Chem.* **2** (4), 274-279 (2010).
15. M. Boero, M. Parrinello, K. Terakura, T. Ikeshoji and C. C. Liew, *Phys. Rev. Lett.* **90** (22), 226403 (2003).
16. P. Wardman, *J. Phys. Chem. Ref. Data* **18** (4), 1637-1755 (1989).
17. S. G. Bratsch, *J. Phys. Chem. Ref. Data* **18** (1), 1-21 (1989).
18. M. S. Pshenichnikov, A. Baltuška and D. A. Wiersma, *Chem. Phys. Lett.* **389** (1), 171-175 (2004).
19. B. C. Garrett, D. A. Dixon, D. M. Camaioni, D. M. Chipman, M. A. Johnson, C. D. Jonah, G. A. Kimmel, J. H. Miller, T. N. Rescigno, P. J. Rossky, S. S. Xantheas, S. D. Colson, A. H. Laufer, D. Ray, P. F. Barbara, D. M. Bartels, K. H. Becker, K. H. Bowen, Jr., S. E. Bradforth, I. Carmichael, J. V. Coe, L. R. Corrales, J. P. Cowin, M. Dupuis, K. B. Eisenthal, J. A. Franz, M. S. Gutowski, K. D. Jordan, B. D. Kay, J. A. LaVerne, S. V. Lymar, T. E. Madey, C. W. McCurdy, D. Meisel, S. Mukamel, A. R. Nilsson, T. M. Orlando, N. G. Petrik, S. M. Pimblott, J. R. Rustad, G. K. Schenter, S. J. Singer, A. Tokmakoff, L.-S. Wang and T. S. Zwier, *Chem. Rev.* **105** (1), 355-390 (2005).
20. X.-F. Gao, D. J. Hood, X. Zhao and G. M. Nathanson, *J. Am. Chem. Soc.* **145** (20), 10987-10990 (2023).
21. B. F. Bachman, D. Zhu, J. Bandy, L. Zhang and R. J. Hamers, *ACS Meas. Sci. Au* **2** (1), 46-56 (2022).
22. H. Lu, F. H. Long, R. M. Bowman and K. B. Eisenthal, *J. Phys. Chem.* **93** (1), 27-28 (1989).
23. S. Kratz, J. Torres-Alacan, J. Urbanek, J. Lindner and P. Vöhringer, *Phys. Chem. Chem. Phys.* **12** (38), 12169-12176 (2010).
24. J. Torres-Alacan, S. Kratz and P. Vöhringer, *Phys. Chem. Chem. Phys.* **13** (46), 20806-20819 (2011).
25. C. L. Thomsen, D. Madsen, S. R. Keiding, J. Tho/gersen and O. Christiansen, *J. Chem. Phys.* **110** (7), 3453-3462 (1999).
26. T. Goulet and J. P. Jay‐Gerin, *J. Chem. Phys.* **96** (7), 5076-5087 (1992).
27. J. L. McGowen, H. M. Ajo, J. Z. Zhang and B. J. Schwartz, *Chem. Phys. Lett.* **231** (4), 504-510 (1994).
28. P. Rumbach, D. M. Bartels, R. M. Sankaran and D. B. Go, *Nat Commun.* **6** (1), 7248 (2015).

29. Y. Inagaki and K. Sasaki, *Plasma Sources Sci. Technol.* **31** (3), 03LT02 (2022).
30. P. Rumbach, D. M. Bartels, R. M. Sankaran and D. B. Go, *J. Phys. D: Appl. Phys.* **48** (42), 424001 (2015).
31. Y. Inagaki and K. Sasaki, *Plasma Sources Sci. Technol.* **32** (6), 065019 (2023).
32. D. T. Elg, H. E. Delgado, D. C. Martin, R. M. Sankaran, P. Rumbach, D. M. Bartels and D. B. Go, *Spectrochim. Acta, Part B* **186**, 106307 (2021).
33. P. J. Bruggeman, A. Bogaerts, J. M. Pouvesle, E. Robert and E. J. Szili, *J. App. Phys.* **130** (20) (2021).
34. P. J. Bruggeman, M. J. Kushner, B. R. Locke, J. G. E. Gardeniers, W. G. Graham, D. B. Graves, R. C. H. M. Hofman-Caris, D. Maric, J. P. Reid, E. Ceriani, D. Fernandez Rivas, J. E. Foster, S. C. Garrick, Y. Gorbanev, S. Hamaguchi, F. Iza, H. Jablonowski, E. Klimova, J. Kolb, F. Krcma, P. Lukes, Z. Machala, I. Marinov, D. Mariotti, S. Mededovic Thagard, D. Minakata, E. C. Neyts, J. Pawlat, Z. L. Petrovic, R. Pflieger, S. Reuter, D. C. Schram, S. Schröter, M. Shiraiwa, B. Tarabová, P. A. Tsai, J. R. R. Verlet, T. von Woedtke, K. R. Wilson, K. Yasui and G. Zvereva, *Plasma Sources Sci. Technol.* **25** (5), 053002 (2016).
35. J. Wang, N. B. Üner, S. E. Dubowsky, M. P. Confer, R. Bhargava, Y. Sun, Y. Zhou, R. M. Sankaran and J. S. Moore, *J. Am. Chem. Soc.* **145** (19), 10470-10474 (2023).
36. N. Sakakibara, T. Ito, K. Terashima, Y. Hakuta and E. Miura, *Phys. Rev. E* **102** (5), 053207 (2020).
37. P. M. Hare, E. A. Price and D. M. Bartels, *J. Phys. Chem. A* **112** (30), 6800-6802 (2008).
38. M. Kalal, O. Slezak, M. Martinkova and Y. J. Rhee, *J. Korean Phys. Soc* **56** (11), 287-294 (2010).
39. C. E. Rallis, T. G. Burwitz, P. R. Andrews, M. Zohrabi, R. Averin, S. De, B. Bergues, B. Jochim, A. V. Voznyuk, N. Gregerson, B. Gaire, I. Znakovskaya, J. McKenna, K. D. Carnes, M. F. Kling, I. Ben-Itzhak and E. Wells, *Rev. Sci. Instrum.* **85** (11) (2014).
40. C. Killer, in *MATLAB Central File Exchange* (2016), Vol. 2023.
41. E. S. Efimenko, Y. A. Malkov, A. A. Murzanev and A. N. Stepanov, *J. Opt. Soc. Am. B* **31** (3), 534-541 (2014).
42. J. Noack and A. Vogel, *IEEE J. Quantum Electron.* **35** (8), 1156-1167 (1999).
43. Z. Yang, C. Zhang, H. Zhang and J. Lu, *Opt. Commun.* **546**, 129803 (2023).
44. E. Abraham, K. Minoshima and H. Matsumoto, *Opt. Commun.* **176** (4), 441-452 (2000).
45. C. B. Schaffer, N. Nishimura, E. N. Glezer, A. M. T. Kim and E. Mazur, *Opt. Express* **10** (3), 196-203 (2002).
46. F. V. Potemkin and E. I. Mareev, *Laser Phys. Lett.* **12** (1), 015405 (2015).
47. D. C. Martin, D. T. Elg, H. E. Delgado, H. M. Nguyen, P. Rumbach, D. M. Bartels and D. B. Go, *Langmuir* **40** (28), 14224-14232 (2024).
48. H. Yui, Y. Yoneda, T. Kitamori and T. Sawada, *Phys. Rev. Lett.* **82** (20), 4110-4113 (1999).
49. Y. Du, E. Price and D. M. Bartels, *Chem. Phys. Lett.* **438** (4), 234-237 (2007).
50. F. Uhlig, J. M. Herbert, M. P. Coons and P. Jungwirth, *J. Phys. Chem. A* **118** (35), 7507-7515 (2014).
51. S. J. Park and B. J. Schwartz, *J. Chem. Theory Comput.* **18** (8), 4973-4982 (2022).
52. J. M. Herbert and M. P. Coons, *Annu. Rev. Phys. Chem.* **68** (Volume 68, 2017), 447-472 (2017).
53. T. Tokushima, Y. Harada, O. Takahashi, Y. Senba, H. Ohashi, L. G. M. Pettersson, A. Nilsson and S. Shin, *Chem. Phys. Lett.* **460** (4), 387-400 (2008).
54. N. Sakakibara, K. Inoue, S. Takahashi, T. Goto, T. Ito, K. Akada, J. Miyawaki, Y. Hakuta, K. Terashima and Y. Harada, *Phys. Chem. Chem. Phys.* **23** (17), 10468-10474 (2021).

# Picosecond imaging of dynamics of solvated electrons during femtosecond laser-induced plasma generation in water

Noritaka Sakakibara,[1,2,3,*] Tsuyohito Ito,[1,2] Yukiya Hakuta,[2] Yoshiki Shimizu,[2] Kazuo Terashima,[1,2] and Eisuke Miura[2,*]

[1] Department of Advanced Materials Science, Graduate School of Frontier Sciences, The University of Tokyo, 5-1-5 Kashiwanoha, Kashiwa, Chiba 277-8561, Japan

[2] AIST-UTokyo Advanced Operando-Measurement Technology Open Innovation Laboratory (OPERANDO-OIL), National Institute of Advanced Industrial Science and Technology (AIST), 5-1-5 Kashiwanoha, Kashiwa, Chiba, 277-8589, Japan

[3] Department of Chemistry, School of Science, Tokyo Institute of Technology, 2-12-1-NE-2 Ookayama, Meguro, Tokyo 152-8550, Japan

---

[*] E-mail: nori.sakakibara@gmail.com

[*] E-mail: e-miura@aist.go.jp

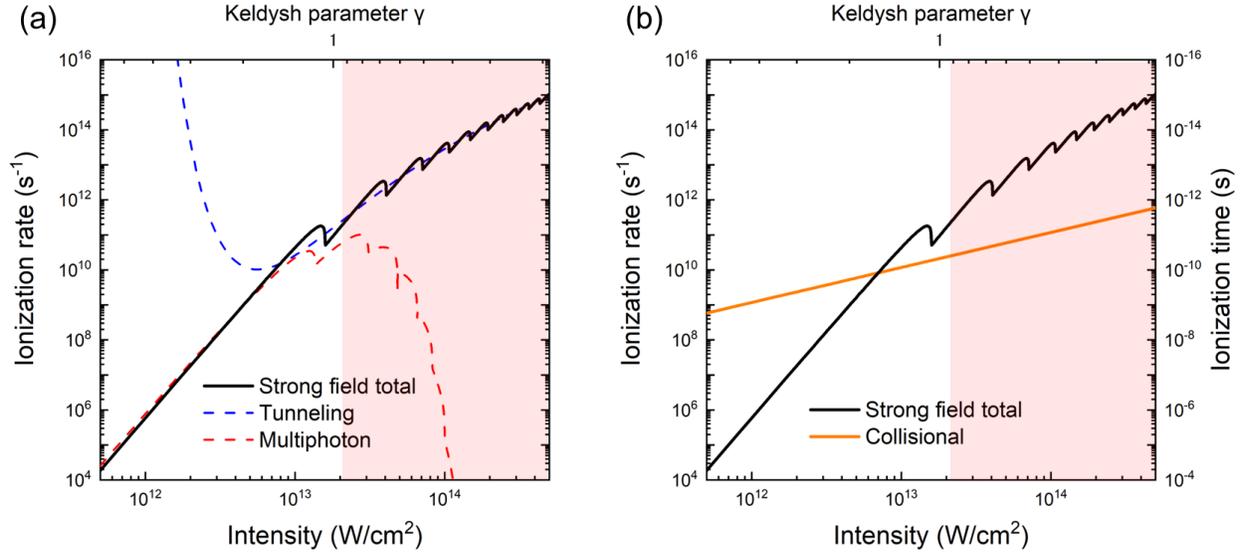

**Figure S1.** Laser intensity dependence of ionization rate in water at 800 nm. (a) Ionization rate by strong electromagnetic field (black solid line) calculated by the Keldysh theory.[1] Two contributions, *i.e.*, tunneling ionization (blue dashed line) and multiphoton ionization (red dashed line), are also shown. (b) Ionization rate by strong electromagnetic field (black solid line) and collisional ionization rate (orange solid line). Collisional ionization rate was calculated based on Drude theory.[1] Corresponding ionization time is also shown in the right axis. In the red-marked area (laser intensity > $2\times10^{13}$ W/cm$^2$), two-step evolution of optical density (OD) was observed.

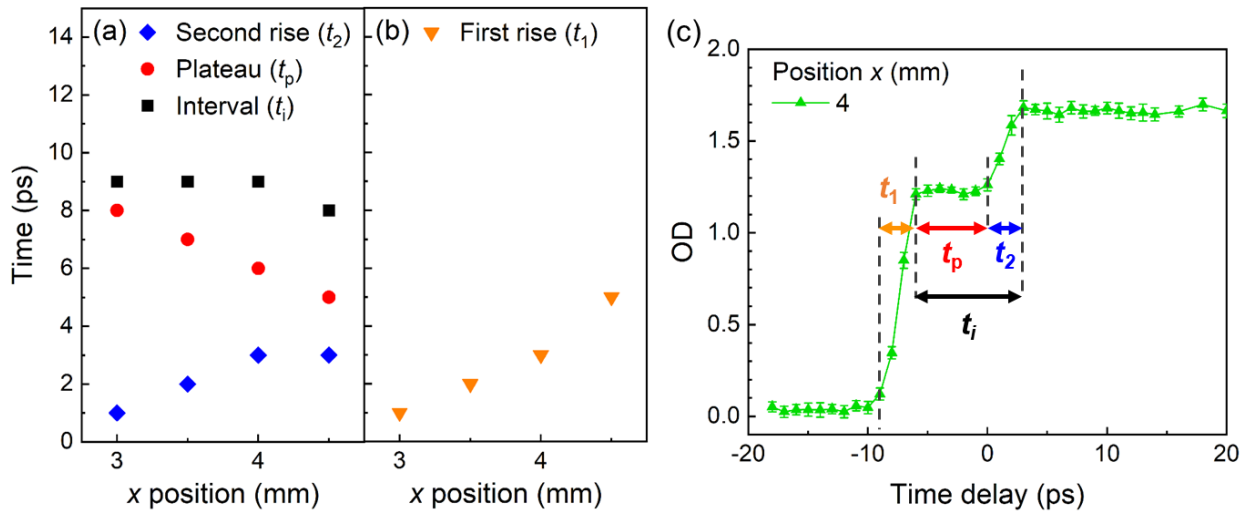

**Figure S2.** (a) Time for the second OD increase ($t_2$, blue diamonds), the duration time of the plateau between the first and second OD increases ($t_p$, red circles) and the interval time between the first and second OD increases ($t_i$, black squares), *i.e.*, the sum of $t_p$ and $t_2$. (b) Time for the first OD increase ($t_1$, orange triangles). (c) Illustration of the times ($t_1$, $t_2$, $t_p$ and $t_i$) in the case of temporal OD behavior at $x$ = 4 mm. The pump pulse energy was 5 mJ. The wavelength of probe pulse was 800 nm. The OD dynamics corresponding to this figure is shown in Figure 2.

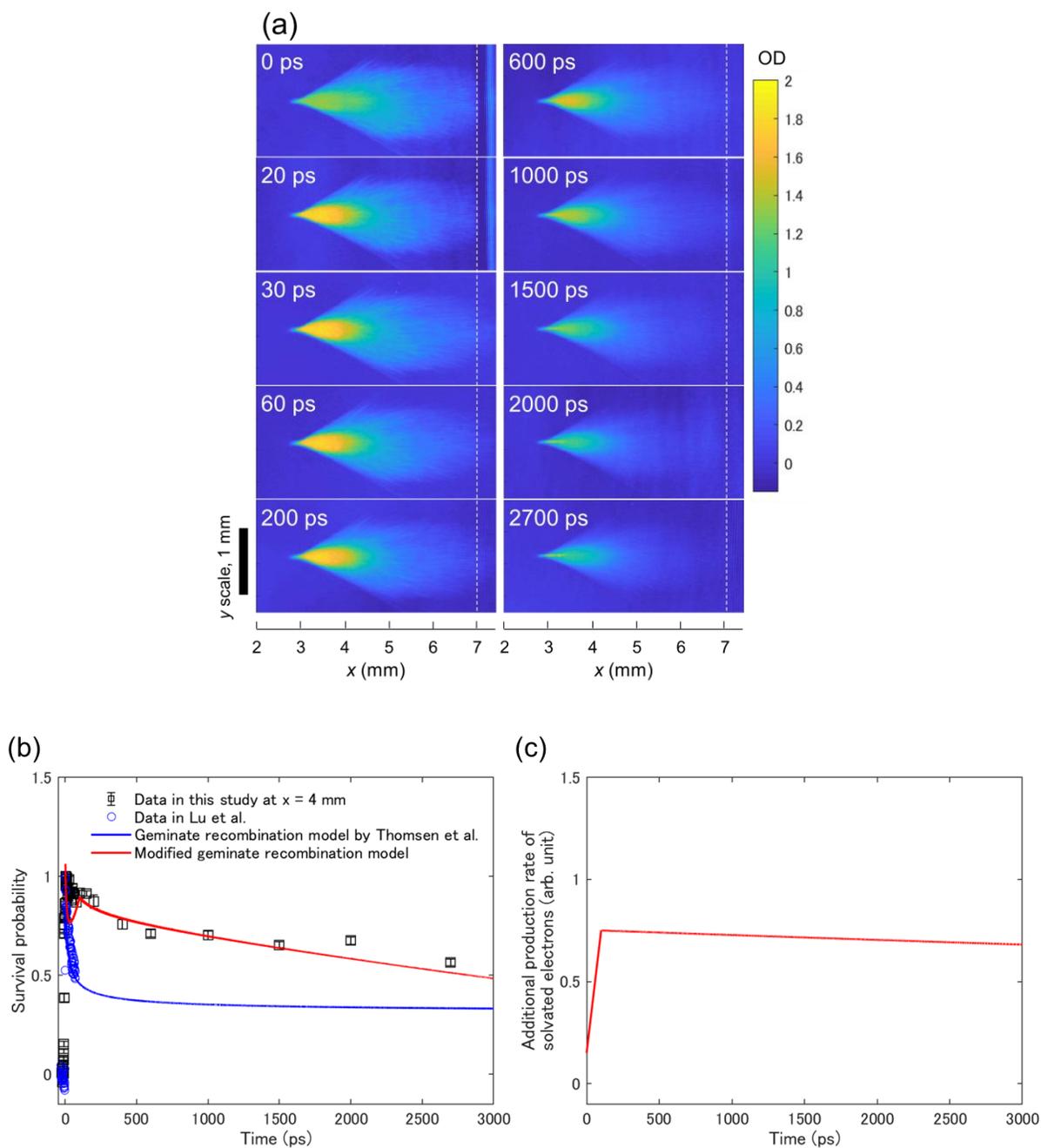

**Figure S3.** (a) Images of OD distribution probed at 800 nm in picosecond to nanosecond time scales after the pump pulse (5 mJ) passed the focal point. The color bar indicates the value of OD. Position $x$ indicates the axial direction along the pump pulse propagation. A pump pulse propagated from the right side of the images. The white dashed lines at the right side of the images indicate the wall of a cuvette. (b) Decay kinetics of OD at $x = 4$ mm (black squares). The decay was fitted by a modified model of geminate recombination which considers additional production of solvated electrons, which was constructed in our previous study.[2] Blue circles are the previously reported decay data of solvated electrons in a photolysis study by Lu et al.[3], which was fitted well by the classical geminate recombination model by Thomsen et al.[4] (c) Normalized rate of additional production of solvated electrons that was assumed in the model calculation.

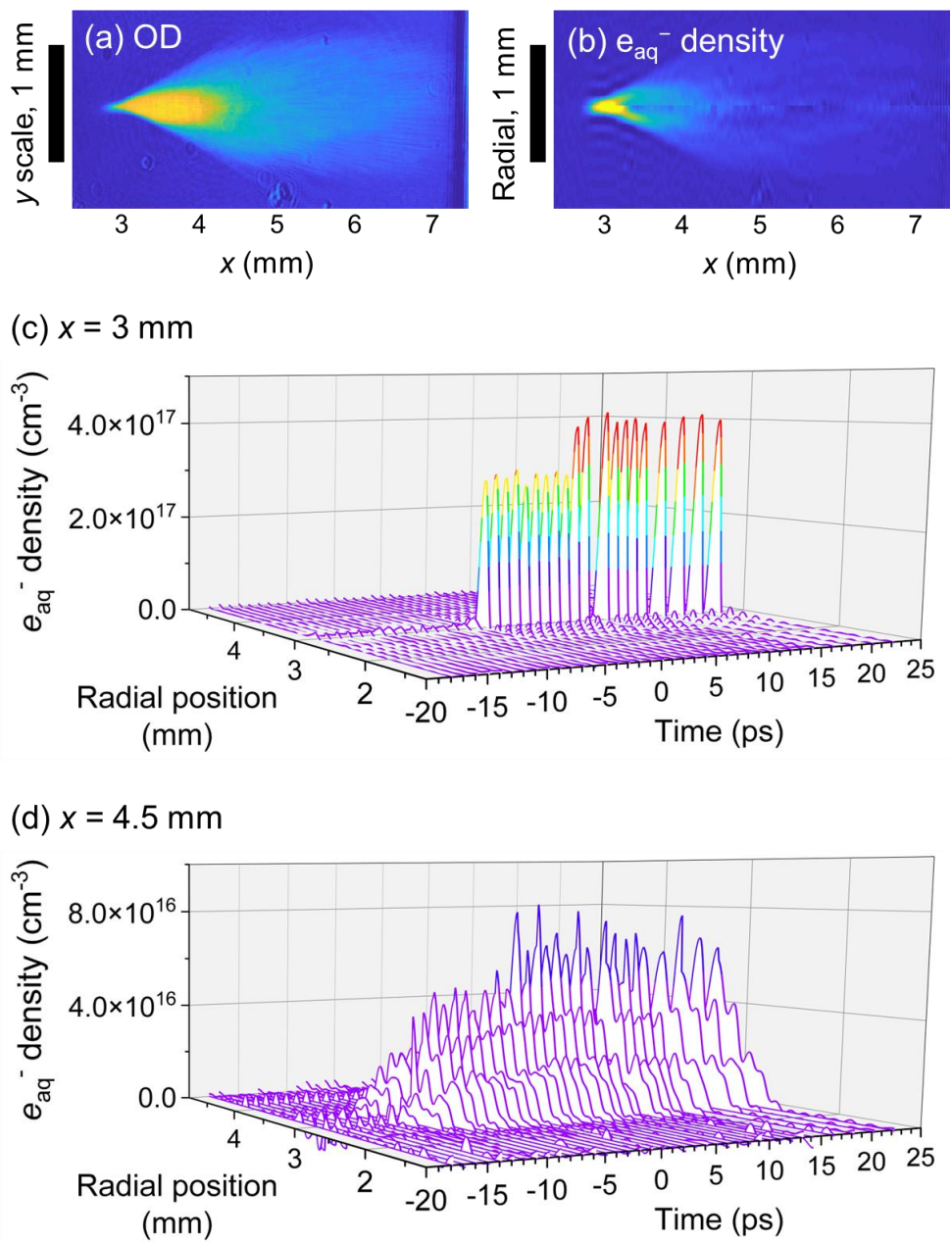

**Figure S4.** Distributions of (a) OD and (b) density of solvated electrons at $t = 20$ ps with 5 mJ pump pulse probed by 800 nm probe pulse. Radial distributions of the density of solvated electrons at (c) $x = 3$ mm and (d) $x = 4.5$ mm are shown with different time delays.

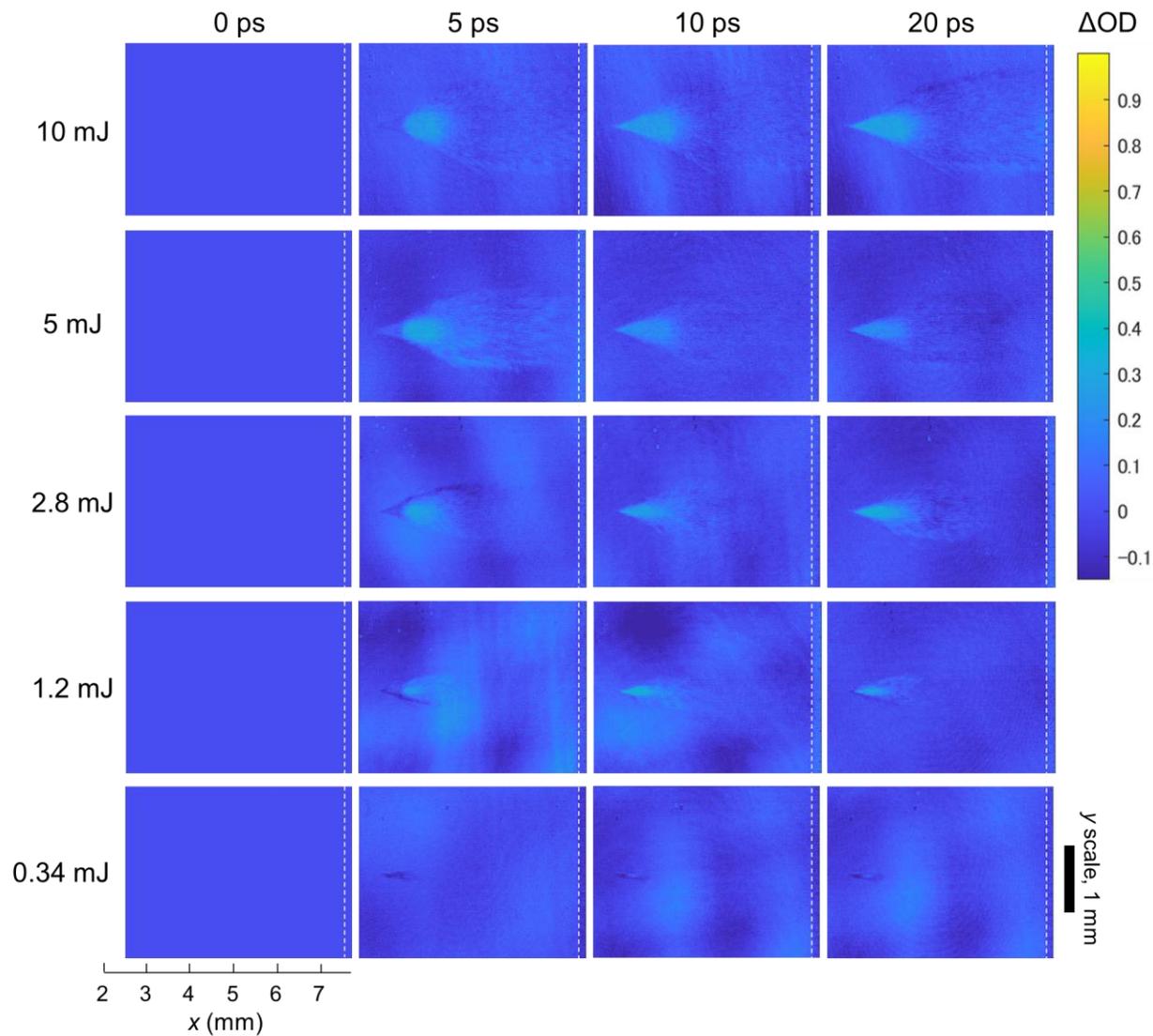

**Figure S5.** Images of a difference of OD values (ΔOD) at $t$ = 0, 5, 10 and 20 ps between $t$ = 0 ps with different pulse energies, corresponding to the picosecond time-resolved OD images in Fig. 4. Further OD increases were observed after 0 ps time delays at every pulse energy used in this study.

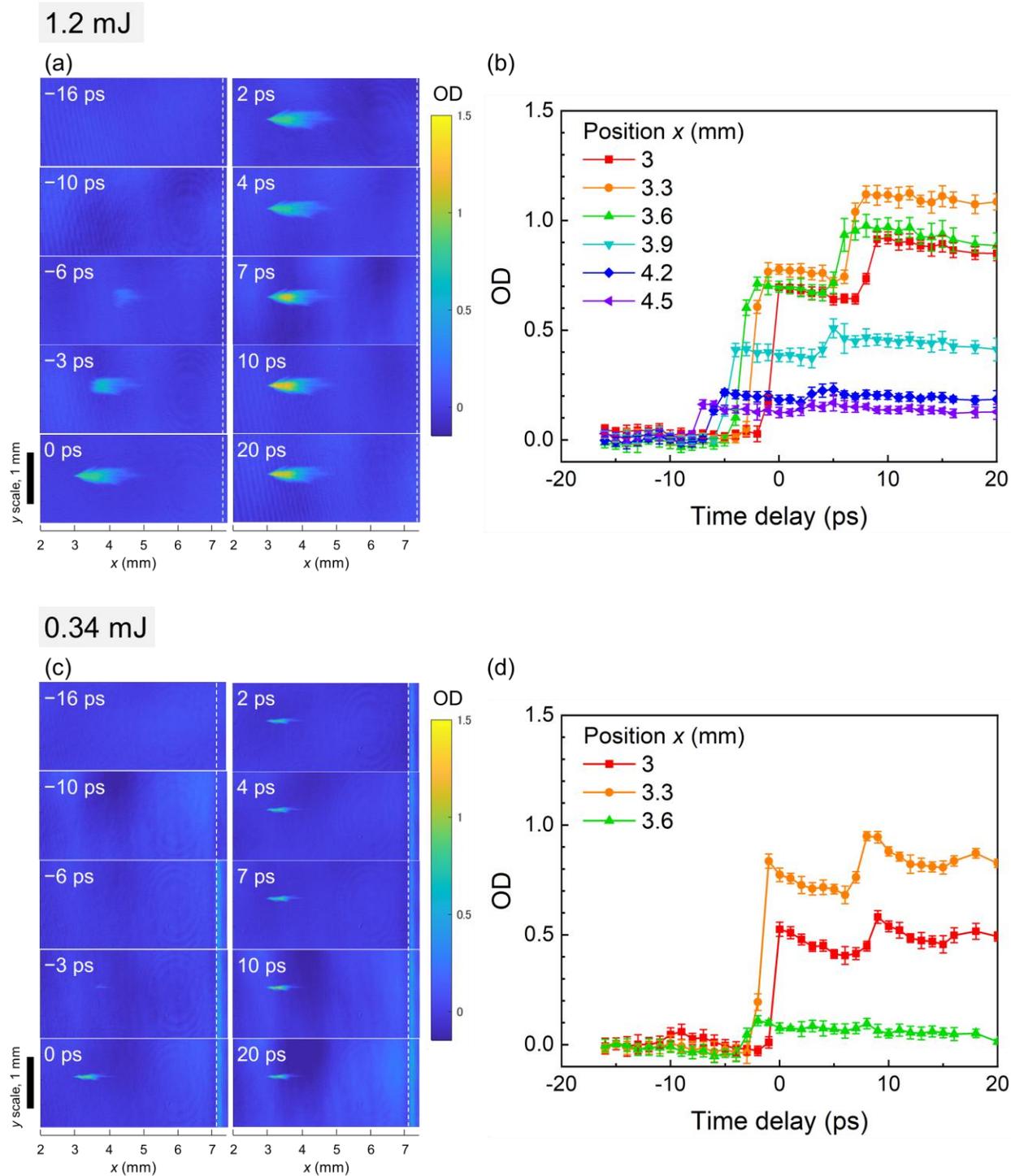

**Figure S6.** Dynamics of solvated electrons probed at 800 nm with 1.2 or 0.34 mJ of pump pulse energies. (a) OD distributions monitored at different time delays during the propagation of the pump pulse (1.2 mJ) and (b) corresponding time evolution of OD at the lateral ($y$-directional) center of the OD distribution with different axial ($x$) positions. (c) OD distributions monitored at different time delays during the propagation of the pump pulse (0.34 mJ) and (d) corresponding time evolution of OD at the $y$-directional center of the OD distribution with different $x$ positions. The pump pulse propagated from the right side of the images.

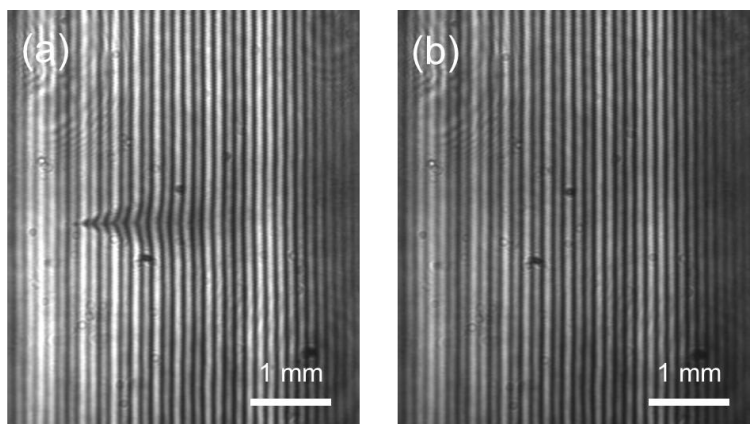

**Figure S7.** Interferometry images of at $t = 0$ ps (a) with and (b) without a pump pulse (5 mJ), probed by 400 nm probe pulse.

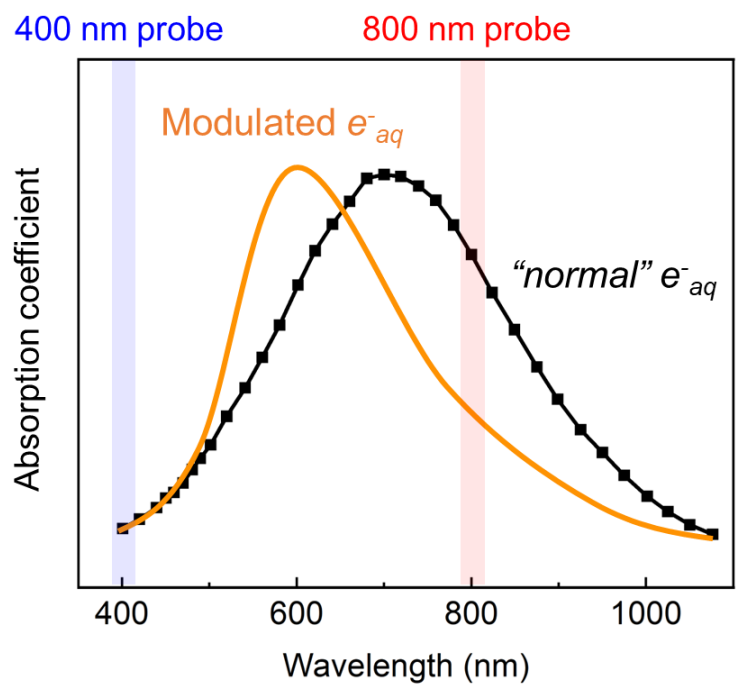

**Figure S8.** An illustration of the hypothesized spectral change of the absorption coefficient of solvated electrons. The spectral change apparently decreases OD value at 800 nm in comparison with the OD of "normal" solvated electron, whereas OD values of modulated and normal solvated electrons are similar at 400 nm.

**References**

1. E. S. Efimenko, Y. A. Malkov, A. A. Murzanev and A. N. Stepanov, *J. Opt. Soc. Am. B* **31** (3), 534-541 (2014).

2. N. Sakakibara, T. Ito, K. Terashima, Y. Hakuta and E. Miura, *Phys. Rev. E* **102** (5), 053207 (2020).

3. H. Lu, F. H. Long, R. M. Bowman and K. B. Eisenthal, *J. Phys. Chem.* **93** (1), 27-28 (1989).

4. C. L. Thomsen, D. Madsen, S. R. Keiding, J. Tho/gersen and O. Christiansen, *J. Chem. Phys.* **110** (7), 3453-3462 (1999).